\documentclass[pra,showpacs,preprintnumbers,amsmath,amssymb,floatfix]{revtex4}
\usepackage{dcolumn}
\usepackage{bm}

\usepackage{amsmath}    
\usepackage{graphicx}   

\def\oper#1{\hat{\mathbf{#1}}} 

\begin{document}

\title{Analytic pulse design for selective population transfer in many-level quantum systems: maximizing the amplitude of population oscillations}
\author{Duje Bonacci}
\affiliation{Physical Chemistry Department, R. Bo\v{s}kovi\'c
Institute, Bijeni\v{c}ka 54, 10000 Zagreb, Croatia}
\email{dbonacci@irb.hr}

\date{\today}

\begin{abstract}

State selective preparation and manipulation of discrete-level
quantum systems such as atoms, molecules or quantum dots is the
ultimate tool for many diverse fields such as laser control of
chemical reactions, atom optics, high-precision metrology and
quantum computing. Rabi oscillations are one of the simplest, yet
potentially quite useful mechanisms for achieving such
manipulation. Rabi theory establishes that in the two-level
systems resonant drive leads to the periodic and complete
population oscillations between the two system levels. In this
paper an analytic optimization algorithm for producing Rabi-like
oscillations in the general discrete many-level quantum system is
presented.

\end{abstract}

\pacs{3.65.Sq}

\maketitle

\newpage
\section{Introduction}

During the past 20 years a number of methods has been devised for
state selective preparation and manipulation of discrete-level
quantum systems
\cite{paramonov1983,chelkowski1990,kaluza1993,bergmann1998,rabitz2003}.
However, simple population oscillations, induced by a resonant
driving pulse have received negligible attention as a prospective
population manipulation method. This might be attributed to two
reasons. The first is that Rabi theory is based on the rotating
wave approximation (RWA), and all attempts to generalize it
without RWA (e.g. \cite{shariar2002.1,barata2000,fujii2003}) are
mathematically very involved. The second is that no attempt has
been made to analytically generalize the original Rabi theory
beyond two-level systems.

In this paper an analytic extension of Rabi theory to transitions
in many-level systems is presented. The aim is to 'design' a
driving pulse of the form:
\begin{equation}
  F(t)= F_{\rm 0} \; m(t) \; \cos{(\omega(t) \; t)}
\label{pulse}
\end{equation}
by establishing analytical optimization relations between its
parameters: maximum pulse amplitude $F_{0}$, pulse envelope shape
$m(t)$, and time dependent carrier frequency $\omega(t)$. The goal
of this enterprize is twofold. The first is to achieve as complete
as possible transfer of population between two selected states of
the system. The second is to make this transfer as rapid as
possible. These two requirements, however, are contradictory;
population transfer can be accelerated by using a more intense
drive, but at the same time a stronger drive increases involvement
of remaining system levels in population dynamics and hence
deteriorates population transfer between a selected pair of
levels.

In this paper it is shown how, for a pulse of arbitrary shape and
duration $m(t)$, the drive frequency can be analytically optimized
to maximize the population transfer amplitude between selected two
levels. In other words, Rabi oscillation theory is re-formulated
for the case of a many-level system driven by an arbitrary
modulated pulse.

\section{Theoretical analysis}

All calculations in this section are done in a system of units in
which $\hbar=1$.

\subsection{Calculation setup}

A quantum system with N discrete stationary levels with energies
$E_i \ (i=1,...,N)$ is considered. The system is driven by a time
dependent perturbation given in Eq. (\ref{pulse}). In the
interaction picture, the dynamics of the system obeys the
Schroedinger equation:
\begin{equation}
  \frac{d}{dt}\mathbf{a}(t)=-i \oper{V}(t) \mathbf{a}(t) ,
\label{schrodinger}
\end{equation}
where $\mathbf{a}(t)$ is a vector of time-dependent expansion
coefficients $a_1(t),..., a_N(t)$. The N$\times$N Matrix
$\oper{V}(t)$ describes interaction between the system and
perturbation. Explicitly, its elements are given by:
\begin{equation}
  V_{ij}(t)= \frac{F_0 \mu_{ij}}{2} m(t)(e^{i s_{ij}
  (\omega(t)-\omega_{ij}) t}+e^{-i s_{ij} (\omega(t)+\omega_{ij}) \; t}).
\end{equation}
$\mu_{ij}$ is transition moment between the i-th and the j-th
levels induced by the perturbation. $s_{ij}=sign(E_i-E_j)$ and
$\omega_{ij}=|E_i-E_j|$ are respectively the sign and the
magnitude of the resonant frequency for the transition between the
i-th and the j-th level.

The aim is to induce population transfer between two arbitrarily
selected levels, designated by $\alpha$ and $\beta$, directly
coupled by the perturbation (i.e. such that $\mu_{\alpha \beta}\ne
0$). To simplify equations, the time variable t is re-scaled to
$\tau$, with transformation between the two given by:
\begin{equation}
  d\tau = \frac{F_0 \mu_{\alpha \beta}}{2} m(t) dt .
\label{definition tau}
\end{equation}
Then with following substitutions:
\begin{eqnarray}
  f_{ij}(\tau)&=&s_{ij} \frac{2}{F_0 \mu_{\alpha \beta}} (\omega(t)-\omega_{ij}) \label{definition f} \\
  g_{ij}(\tau)&=&s_{ij} \frac{2}{F_0 \mu_{\alpha \beta}} (\omega(t)+\omega_{ij}) \label{definition g} \\
  x(\tau)&=& \frac{F_0 \mu_{\alpha \beta}}{2} t(\tau) \label{definition x} \\
  R_{ij} &=& \frac{\mu_{ij}} {\mu_{\alpha \beta}} \label{definition R}
\end{eqnarray}
Eq. (\ref{schrodinger}) transforms into:
\begin{equation}
  \frac{d}{d \tau}\mathbf{a}(\tau)=-i \oper{W}(\tau)\mathbf{a}(\tau),
\label{n level}
\end{equation}
where:
\begin{equation}
  W_{ij}(\tau) = R_{ij}(e^{i f_{ij}(\tau) x(\tau)}+e^{-i g_{ij}(\tau)x(\tau)}).
\label{wovi}
\end{equation}

Initial conditions for the problem of selective population
transfer comprise complete population initially (at $t=\tau=0$)
contained in only one of the selected levels, either $\alpha$ or
$\beta$. The other selected level, as well as all the remaining
N-2 'perturbing' levels of the system are unpopulated at this
time.

Population evolution $\Pi_i(t)$ of the i-th level is determined
from $\Pi_i(t)=|a_i(t)|^2$.

\subsection{Population oscillations in the two-level system - recapitulation}

Having in mind that $s_{ij}=-s_{ji}$, the explicit form of the
general dynamical equation (\ref{n level}) in a two-level system
is:
\begin{eqnarray}
  \frac{d}{d \tau}
    \begin{bmatrix}
      a_\alpha(\tau) \\
      a_\beta(\tau) \\
    \end{bmatrix}
    = -i
    \begin{bmatrix}
    0 & e^{i f_{\alpha \beta}(\tau) x( \tau)} + e^{-i g_{\alpha \beta}(\tau) x(\tau)} \\
    e^{-i f_{\alpha \beta}(\tau) x( \tau)} + e^{i g_{\alpha \beta}(\tau) x(\tau)} & 0 \\
    \end{bmatrix}
    \begin{bmatrix}
      a_\alpha(\tau) \\
      a_\beta(\tau)  \\
    \end{bmatrix}
\label{2 level}
\end{eqnarray}
Under certain conditions (for a thorough discussion see eg.
\cite{bonacci2003.1}) this equation may be simplified by
introducing the rotating wave approximation (RWA). Within the RWA,
dynamical impact of complex exponentials $ e^{\pm g_{\alpha
\beta}(\tau) x(\tau)}$ is neglected and these may be eliminated
from the equation. Hence, (\ref{2 level}) reduces to:
\begin{eqnarray}
  \frac{d}{d \tau}
  \begin{bmatrix}
  a_\alpha(\tau) \\
  a_\beta(\tau) \\
  \end{bmatrix}
  =
  -i
  \begin{bmatrix}
  0 & e^{i f_{\alpha \beta}(\tau) x( \tau)} \\
  e^{-i f_{\alpha \beta}(\tau) x( \tau)} & 0 \\
  \end{bmatrix}
  \begin{bmatrix}
  a_\alpha (\tau) \\
  a_\beta (\tau) \\
  \end{bmatrix}
\label{2 level RWA}
\end{eqnarray}
Finally, adjusting the perturbation frequency to the resonant
value ($\omega(\tau) \equiv \omega_{\alpha \beta}$) results in
$f_{\alpha \beta}(\tau)=0$, and (\ref{2 level RWA}) reduces to:
\begin{equation}
  \frac{d}{d \tau}
  \begin{bmatrix}
  a_\alpha (\tau) \\
  a_\beta(\tau) \\
 \end{bmatrix}
  = -i
  \begin{bmatrix}
  0 & 1 \\
  1 & 0 \\
  \end{bmatrix}
  \begin{bmatrix}
  a_\alpha (\tau) \\
  a_\beta (\tau) \\
  \end{bmatrix}
\label{2 level RWA resonant}
\end{equation}

As may be easily demonstrated by solving this simple equation, the
resonant perturbation induces complete periodic transfer of the
population between two levels. The time $\Theta$ (in units of
$\tau$) or T (in units of t) required for a single population
oscillation is determined from:
\begin{equation}
  \int_0^\Theta d\tau \equiv \frac{F_0 \mu_{\alpha \beta}}{2} \int_0^T m(t)dt = \pi
\label{pi pulse}
\end{equation}
This is a well known result which forms the basis of $\pi$-pulse
theory.

\subsection{Population oscillations in the three-level system}

As will be shown in subsequent sections, the whole analytical
approach to the maximization of the population oscillation
amplitude in a general many-level quantum system may be reduced to
discussion of a three-level system. Along with two 'selected'
levels $\alpha$ and $\beta$, the system now discussed contains one
additional 'perturbing' level, designated with index p. The only
requirements on the system internal structure are that
$\mu_{\alpha \beta}, \mu_{\beta p} \ne 0$ and $\mu_{\alpha p}=0$.
While the first two requirements are necessary, the last one does
not reduce the generality of the final results to any significant
extent and it is introduced for calculational convenience.

The three-level version of the dynamical equation (\ref{n level})
is:

\begin{eqnarray}
\label{3 level}
  \frac{d}{d \tau}
  \begin{bmatrix}
  a_\alpha(\tau) \\
  a_\beta(\tau) \\
  a_p(\tau) \\
  \end{bmatrix}
  &=& \\
  -&i&
  \begin{bmatrix}
  0 & e^{i f_{\alpha \beta}(\tau) x(\tau)} + e^{-i g_{\alpha \beta}(\tau) x(\tau)} & 0 \\
  e^{-i f_{\alpha \beta}(\tau) x(\tau)} + e^{i g_{\alpha \beta}(\tau) x(\tau)} &  0
  & R_{\beta p}(e^{-i f_{\beta p}(\tau) x(\tau)} + e^{i g_{\beta p}(\tau) x(\tau)}) \\
  0 & R_{\beta p}(e^{i f_{\beta p}(\tau) x(\tau)} + e^{-i g_{\beta p}(\tau) x(\tau)}) & 0 \\
  \end{bmatrix}
  \begin{bmatrix}
  a_\alpha(\tau) \\
  a_\beta (\tau) \\
  a_p(\tau) \\
  \end{bmatrix} \nonumber.
\end{eqnarray}

\subsubsection{The adiabatic approach and perturbing level population dynamics}

In order to solve equation (\ref{3 level}) two assumptions shall
be made. The legitimacy of both of them may be checked
retrospectively from the final solution. The first one is that the
RWA may be applied for transition $\alpha \leftrightarrow \beta$
so that exponentials containing $g_{\alpha \beta}(\tau)$ may be
dropped from equation (\ref{3 level}). A simple way to verify this
assumption's legitimacy may be done using Rabi profile plots, as
described in \cite{bonacci2003.1}. The second is that the
dynamical time-scale of solutions for $a_{\alpha,\beta} (\tau)$ is
much longer than that for $a_p(\tau)$. This enables one to regard
$a_\beta(\tau)$ as a slowly changing parameter in the dynamical
equation for $a_p(\tau)$. The solution to this equation hence may
be obtained in terms of a parameter whose value needs not be known
beforehand.

Introducing the first of the assumptions into (\ref{3 level}) and
reformulating it slightly, a set of two coupled differential
equations is obtained:

\begin{eqnarray}
  \frac{d}{d \tau}
  \begin{bmatrix}
  a_\alpha(\tau) \\
  a_\beta(\tau) \\
  \end{bmatrix}
  &=& -i
  \begin{bmatrix}
  0 & e^{i f_{\alpha \beta}(\tau) x( \tau)} \\
  e^{-i f_{\alpha \beta}(\tau) x( \tau)}& 0 \\
  \end{bmatrix}
  \begin{bmatrix}
  a_\alpha(\tau) \\
  a_\beta (\tau) \\
  \end{bmatrix}
  \label{3 level RWA ab}
  - i R_{\beta p} (e^{-i f_{\beta p}(\tau) x(\tau)} + e^{i g_{\beta p}(\tau) x(\tau)})
  \begin{bmatrix}
  0 \\
  1 \\
  \end{bmatrix}
  a_p(\tau) , \\
  \frac{d}{d \tau} a_p(\tau) &=&- i R_{\beta p} (e^{i f_{\beta p}(\tau) x( \tau)} +
e^{-i g_{\beta p}(\tau) x(\tau)})  a_\beta(\tau). \label{3 level
p}
\end{eqnarray}

Now consider just equation (\ref{3 level p}). The formal solution
to this equation is:
\begin{equation}
\label{formal sol}
  a_{p}(\tau)= - i R_{\beta p} \int_{0}^{\tau} (e^{i f_{\beta p}(\tau') x( \tau')} + e^{-i g_{\beta
  p}(\tau') x(\tau')}) a_{\beta } (\tau' ) d\tau'
\end{equation}
This integral cannot be precisely evaluated until the exact form
of the solution $a_{\beta} (\tau)$ and optimized perturbation
frequency $\omega(\tau)$ - through which $f_{\beta p}(\tau)$ and
$g_{\beta p} (\tau)$ are defined ((\ref{definition f}),
(\ref{definition g})) - are known. Certainly, these are not known
before the final solution of the whole optimization procedure is
obtained. However, introduction of the second assumption enables
one to evaluate the partial contribution to the integral in
(\ref{formal sol}) from some interval $ \tau_0<\tau'<\tau $ within
which changes in all these functions are so insignificant that
functions themselves may be approximated by a constant value.

The aim of the optimization procedure is effectively to eliminate
the dynamical impact of the perturbing level on population
transfer between the two selected levels. If this is achieved,
then dynamics of subsystem $(\alpha,\beta)$ will be very similar
to the dynamics of the pure two level system. Hence the population
oscillation period will be about $\Theta=\pi$, which is then the
dynamical time-scale for $a_{\alpha,\beta} (\tau)$. As optimizing
variations of the driving frequency $\omega(\tau)$ are caused
exclusively by the changes in perturbation envelope amplitude
$m(t(\tau))$ it is transparent that dynamical time-scales for
$f_{\beta p}(\tau)$ and $g_{\beta p} (\tau)$ are of the same order
as that for $m(t(\tau))$. The dynamical time-scale of $m(t(\tau))$
must be of the same order as that of $a_{\alpha,\beta} (\tau)$ or
longer, for otherwise not even a single complete population
transfer would be achieved. Hence, in the interval $
\tau_0<\tau'<\tau $ such that $\tau -\tau_0<<\pi$, the optimized
functions $a_\beta (\tau')$, $f_{\beta p}(\tau')$ and $g_{\beta p}
(\tau')$ may be considered constant, $a_{\beta} (\tau') \approx
a_{\beta} (\tau_0)$, $f_{\beta p}(\tau') \approx f_{\beta
p}(\tau)$ and $g_{\beta p} (\tau') \approx g_{\beta p} (\tau_0)$.
Finally, in this interval $x(\tau')$ may be approximated by
$x(\tau') \approx x(\tau_0)+\frac{1}{m(t(\tau_0))} (\tau' -\tau
_0)$. In several simple steps the following result is obtained:
\begin{equation}
\label{apert}
  a_p(\tau) \approx c(\tau_0) - s_{\beta p}\; \sigma_{\beta p}
  \frac{m(t(\tau))}{1-\Delta_{\beta p}(\tau)} \Bigl (1 - \delta_{\beta p}
  \frac{1-\Delta_{\beta p}(\tau)}{1-\delta_{\beta p} \Delta_{\beta p}(\tau)}
  \; e^{-2 i s_{\beta p} \omega (\tau) t(\tau)} \Bigr ) \; e^{ i f_{\beta
  p}(\tau) x(\tau)}\; a_{\beta} (\tau)
\end{equation}
where:
\begin{eqnarray}
  \sigma_{\beta p} &=& \frac{ F_0 \mu_{\beta p}}{2(\omega_{\alpha \beta}
  - \omega_{\beta p})} \ , \label{sigma} \\
  \delta_{\beta p} &=& {\frac{\omega_{\alpha \beta} - \omega_{\beta
  p}}{\omega_{\alpha \beta} + \omega_{\beta p}} } \ ,\label{delta} \\
  \Delta_{\beta p}(\tau) &=& \frac {\omega(\tau)- \omega_{\alpha \beta}}
{\omega_{\beta p} - \omega_{\alpha \beta}} \ .\label{Delta}
\end{eqnarray}
The constant term $c(\tau_0)$ includes both the $a_p(\tau_0)$ and
the integration constant obtained by inserting the lower limit
value $\tau_0$ into the solution. Its exact value cannot be
determined since integration cannot be analytically stretched over
the whole interval from $\tau=0$ to $\tau=\Theta$. However, since
$a_p(\tau=0)=0$ this constant is of the order of average value of
remaining rapid complexly rotating expression, which is very
nearly equal to zero.

If detuning from resonance is assumed small, the perturbation
frequency may be approximated by $\omega(\tau) \approx
\omega_{\alpha \beta}$ so that $\Delta_{\beta p}(\tau)=0$. The
obtained solution (\ref{apert}) then immediately provides an upper
limit for the magnitude of perturbing level population:
\begin{equation}
\label{leak}
  |\Pi_p(\tau)| \leq \sigma_{\beta p}^2 (1+|\delta_{\beta p}|)^2 .
\end{equation}
In general $\delta_{\beta p}<<1$, so the whole bracket can be
reduced to 1, and the maximum amplitude of the perturbing level's
population is roughly $\sigma_{\beta p}^2$. Hence, this quantity
may be regarded as a parameter determining the effective strength
of the perturbation applied to a particular transition: if
$\sigma_{\beta p}^2<<1$, then the dynamical impact of level p is
negligible and the perturbation may be considered weak; if
$\sigma_{\beta p}^2\sim 1$ , the perturbation is very strong. This
result may also be cast into a convenient quantitative form: to
keep the 'leakage' of the population to the perturbing level $p$
below a certain limiting value $M_p$, the greatest drive intensity
which may be employed is roughly:
\begin{equation}
\label{max drive}
  F_0^{max} = |\frac{2(\omega_{\alpha \beta} - \omega_{\beta p})}
  {\mu_{\beta p}}|\sqrt{M_p} .
\end{equation}

\subsubsection{Optimized driving frequency}

When solution (\ref{apert}) is plugged into Eq. (\ref{3 level RWA
ab}) a single closed dynamical equation is obtained for a
two-level sub-system $(\alpha,\beta)$:
\begin{equation}
\label{3 level final ab}
  \frac{d}{d \tau}
  \begin{bmatrix}
  a_\alpha(\tau) \\
  a_\beta(\tau) \\
  \end{bmatrix}
   =  - i
  \begin{bmatrix}
  0 & e^{i f_{\alpha \beta}(\tau) x( \tau)} \\
  e^{-i f_{\alpha \beta}(\tau) x( \tau)} & -\chi (\tau) \\
  \end{bmatrix}
  \begin{bmatrix}
  a_\alpha(\tau) \\
  a_\beta (\tau) \\
  \end{bmatrix},
\end{equation}
with:
\begin{equation}
\label{def chi}
  \chi (\tau ) =  s_{\beta p}\;
  \sigma_{\beta p} \; R_{\beta p} \frac{m(t(\tau))}
  {1-\Delta_{\beta p}(\tau)} (1 - \delta_{\beta p}
  \frac{1-\Delta_{\beta p}(\tau)}{1-\delta_{\beta p}
  \Delta_{\beta p}(\tau)} \; e^{-2 i s_{\beta p} \omega (\tau) t(\tau)} )
  (1 + e^{2 i s_{\beta p} \omega (\tau) t(\tau)} )
\end{equation}

Now a transformation of the $(\alpha,\beta)$ sub-system vector is
sought:
\begin{equation}
\label{transformation}
  \begin{bmatrix}
  b_\alpha(\tau) \\
  b_\beta(\tau) \\
  \end{bmatrix}
  = e^{-i \oper{\Lambda} (\tau)}
  \begin{bmatrix}
  a_\alpha(\tau) \\
  a_\beta(\tau) \\
  \end{bmatrix}
\end{equation}
with:
\begin{eqnarray}
  \oper{\Lambda} (\tau) =
  \begin{bmatrix}
  \rho_1 (\tau) & 0 \\
  0 & \rho _2(\tau) \\
  \end{bmatrix}.
\end{eqnarray}
such that the transformed sub-system vector satisfies:
\begin{equation}
\label{transformed to rabi}
  \frac {d }{d \tau}
  \begin{bmatrix}
  b_\alpha(\tau) \cr
  b_\beta(\tau)
  \end{bmatrix}
  = - i
  \begin{bmatrix}
  0 & 1 \cr
  1 & 0 \cr
  \end{bmatrix}
  \begin{bmatrix}
  b_\alpha(\tau) \cr
  b_\beta(\tau)
  \end{bmatrix}
\end{equation}
As this equation is identical to (\ref{2 level RWA resonant}), the
corresponding solution would represent complete population
transfer oscillations between levels $\alpha$ and $\beta$.
Introducing transformation (\ref{transformation}) into (\ref{3
level final ab}), the following equation is obtained:
\begin{equation}
  \frac {d }{d \tau }
  \begin{bmatrix}
  b_\alpha(\tau) \\
  b_\beta(\tau) \\
  \end{bmatrix}
  = - i
  \begin{bmatrix}
  \frac{d}{d\tau}\rho _1(\tau) & e^{i (f_{\alpha \beta}(\tau)
  x(\tau) - (\rho _1 (\tau)- \rho _2(\tau)))}  \\
  e^{-i (f_{\alpha \beta}(\tau) x(\tau) - (\rho _1 (\tau)- \rho _2(\tau))) }  & -\chi
  (\tau)+ \frac{d}{d\tau} {\rho _2}(\tau) \\
  \end{bmatrix}
  \begin{bmatrix}
  b_\alpha(\tau) \\
  b_\beta(\tau) \\
  \end{bmatrix}
\end{equation}
If this is to be fitted to form (\ref{transformed to rabi}), the
following conditions must be fulfilled:
\begin{eqnarray}
  \frac{d}{d\tau} {\rho _1}(\tau)&=& 0 , \\
  \chi (\tau ) - \frac{d}{d\tau}{\rho _2}(\tau)&=& 0, \\
  f_{\alpha \beta}(\tau) x(\tau) - (\rho _1 (\tau)- \rho_2(\tau))&=& 0 ,
\end{eqnarray}
which can be compactly written as:
\begin{equation}
\label{difjed}
  \frac{d}{d \tau} (f_{\alpha \beta}(\tau) x(\tau))=-\chi (\tau)
\end{equation}
Integration of the last equation yields the formal solution for
$\Delta_{\beta p}(\tau)$ from which optimized perturbation
frequency $\omega(\tau)$ may be extracted:
\begin{eqnarray}
  \Delta_{\beta p}(\tau)&=&(s_{\beta \alpha} s_{\beta p})
  \sigma_{\beta p}^2 \frac{1}{x(\tau)}  \int_0^\tau
  \frac{1- \delta_{\beta p} }{(1-\Delta_{\beta p}(\tau'))
  (1-\delta_{\beta p} \Delta_{\beta p}(\tau'))} m(t(\tau))
  d\tau' \label{Delta solution} \\
  &+& (s_{\beta \alpha} s_{\beta p})\sigma_{\beta p}^2
  \frac{1}{x(\tau)} \int_0^\tau \left( \frac{e^{2 i s_{\beta p}
  \omega (\tau') t(\tau')} }{1-\Delta_{\beta p}(\tau')} -
  \delta_{\beta p} \frac{ e^{-2 i s_{\beta p} \omega (\tau')
  t(\tau')} }{1-\delta_{\beta p} \Delta_{\beta p}(\tau')}\right)
  m(t(\tau)) d\tau' \nonumber
\end{eqnarray}
Since $\Delta_{\beta p}(\tau)$ is generally a small quantity, this
equation may be solved iteratively, using $\Delta_{\beta
p}(\tau')=0$ as the initial value. The contribution to the total
$\Delta_{\beta p}(\tau)$ from the second integral, containing
rapidly rotating complex exponentials, may be shown to be minor
compared to the one from the first, real integral. However, the
very fact that there is an imaginary contribution to the optimized
perturbation frequency indicates that the optimization procedure
simply cannot completely annihilate the dynamical impact of the
perturbing level. Nevertheless, as will be demonstrated in the
following section, it may be done to a very good approximation.

Transforming (\ref{Delta solution}) back to the original time
coordinate $t$ and keeping only the first integral yields the
approximate recurrent solution for $\Delta_{\beta p}(t)$:
\begin{equation}
  \Delta_{\beta p}(t)=(s_{\beta \alpha} s_{\beta p})\sigma_{\beta p}^2
  (1- \delta_{\beta p}) \frac{1}{t} \int_0^t \frac{(m(t'))^2}
  {(1-\Delta_{\beta p}(t'))(1-\delta_{\beta p} \Delta_{\beta p}(t'))} dt'
\label{Delta recurrent}
\end{equation}
Introducing the zeroth-order approximation $\Delta_{\beta
p}(\tau')=0$ into this equation, the analytic expression for the
first-order optimized frequency is obtained:
\begin{equation}
  \omega (t) = \omega_{\alpha \beta}+(\omega_{\beta
  p}-\omega_{\alpha \beta}) (s_{ \beta\alpha} s_{\beta p})
  \sigma_{\beta p}^2 (1-\delta_{\beta p}) \frac{1}{t} \int_0^t
  (m(t'))^2 dt'
\label{bez rot}
\end{equation}
For all but the strongest perturbations (i.e. such that
$\sigma_{\beta p}^2 \sim 1$ or greater) higher-order corrections
are not needed.

Note that the frequency shift in (\ref{Delta recurrent}) and
(\ref{bez rot}) may be either away from the perturbing line or
towards it. Which case it will be depends on the relation between
the energies of the three system levels: if level $\beta$ has
either the highest or the lowest total energy, so that both
transitions $\beta \rightarrow \alpha$ and $\beta \rightarrow p$
are energy-wise either 'downwards' ($s_{ \beta\alpha}=s_{\beta
p}=+1$) or 'upwards' ($s_{ \beta\alpha}=s_{\beta p}=-1$) then the
shift will be \textit{towards} the perturbing line; on the other
hand, if the energy of level $\beta$ is in between the other two
energies so that these two transitions are in the opposite
directions ($s_{ \beta\alpha}=-s_{\beta p}$), the shift will be
\textit{away} from the perturbing line.

\subsection{Population oscillations in the many-level system}

The approach presented in Sec. II.3 can be easily extended to
include multiple perturbation levels. The perturbation now couples
each of the selected levels to a certain number of perturbing
levels, and each of perturbing levels' dynamics is calculated
independently from all the others. This may be done as long as the
perturbation due to each single perturbing level is kept
reasonably small (gauged by standards of the three-level system).
Accordingly, the many-level analog of Eq. (\ref{3 level final ab})
is:
\begin{equation}
  \frac {d}{d \tau}
  \begin{bmatrix}
  a_\alpha(\tau) \\
  a_\beta(\tau) \\
  \end{bmatrix}
   = - i
  \begin{bmatrix}
   -\chi _{\alpha}(\tau) & e^{i f_{\alpha \beta}(\tau ) x(\tau)} \\
   e^{-i f_{\alpha \beta}(\tau ) x(\tau)}  & -\chi_{\beta} (\tau) \\
  \end{bmatrix}
  \begin{bmatrix}
   a_\alpha(\tau) \\
   a_\beta(\tau) \\
  \end{bmatrix}
\label{2multi}
\end{equation}
with:
\begin{eqnarray}
  \chi_{\alpha} (\tau ) &=&  m(t(\tau))\sum_{q=1}^n s_{\alpha q}\;
  \sigma_{\alpha q} R_{\alpha q} \; \frac{1 - \delta_{\alpha
  q}}{(1-\Delta_{\alpha q}(\tau))(1-\delta_{\alpha q} \Delta_{\alpha q}(\tau))}, \\
  \chi_{\beta} (\tau ) &=&  m(t(\tau))\sum_{p=1}^n s_{\beta p}\;
  \sigma_{\beta p} R_{\beta p} \; \frac{1 - \delta_{\beta p}}
  {(1-\Delta_{\beta p}(\tau))(1-\delta_{\beta p} \Delta_{\beta p}(\tau))} ,
\end{eqnarray}
where all quantities are defined analogously to the ones in the
previous section, and complex rotating terms have been eliminated.

The many-level equivalent of (\ref{difjed}) is:
\begin{equation}
  \frac{d}{d\tau} \Bigl ( f_{\alpha \beta}(\tau) x(\tau) \Bigr )
  =\Bigl ( \chi_{\alpha} (\tau) - \chi _{\beta} (\tau ) \Bigr ),
\end{equation}
and the first-order solution is:
\begin{eqnarray}
  \omega (t) = \omega_{\alpha \beta} + \Bigl (&\sum_{q=1}^m&
  (\omega_{\alpha q}-\omega_{\alpha \beta}) (s_{\alpha \beta}
  s_{\alpha q}) \sigma_{\alpha q}^2(1-\delta_{\alpha q}) \label{bez rot multi} \\
  + &\sum_{p=1}^n& (\omega_{\beta p}-\omega_{\alpha \beta})(s_{\beta
  \alpha} s_{\beta p}) \sigma_{\beta p}^2(1-\delta_{\beta p})\Bigr )
  \frac{1}{t} \int_0^t (m(t'))^2 dt' \nonumber.
\end{eqnarray}

The total perturbation intensity for the case of a many-level
system may be estimated by considering the maximum total
population of all perturbing levels:
\begin{eqnarray}
  \sigma_{tot}^2 &\equiv& \sum_{q=1}^m Max(\Pi_q(t;T_0<t<T))+
  \sum_{p=1}^n Max(\Pi_q(t;T_0<t<T)) \nonumber \\ &=& \sum_{q=1}^m
  \sigma_{\alpha q}^2 + \sum_{p=1}^n \sigma_{\beta p}^2
  \label{sigma tot}
\end{eqnarray}
Hence, if $\sigma_{tot}^2<<1$ the perturbation of the many-level
system is small; otherwise it is large. In the following section
it shall be demonstrated that (\ref{sigma tot}) provides not just
a qualitative, but also an excellent quantitative criterion for
determination of the impact of perturbing levels on population
oscillation dynamics.

\section{Numerical simulations}

In this section numerical simulations of system dynamics for
resonant (i.e. un-optimized) and optimized (determined from
(\ref{Delta recurrent}) and (\ref{bez rot multi})) perturbation
frequencies are presented and compared. Several pulse envelope
shapes are considered: square pulse ($m(t)=1$), sine pulse
($m(t)=sin(\Omega t)$) and sine squared pulse ($m(t)=(sin(\Omega
t))^2$) (Fig. 1).

\begin{figure}
  \includegraphics[width=10cm]{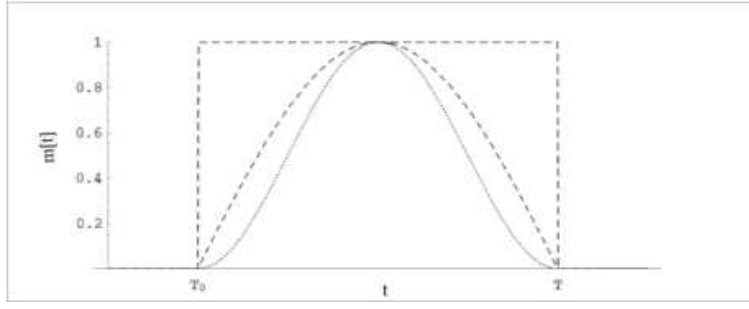} \\
  \caption{Pulse envelope profiles. Time t is on the abscissa and
  $m(t)$ is on the ordinate. The dash-dotted line corresponds to a square pulse,
the dashed line to a sine pulse and the dotted line to a sine
squared pulse. The pulse is switched on at $t=T_0$ and lasts until
  $t=T$. In all three cases, the maximum value of the perturbing
  field intensity field, $F_0$, achieved at time $\frac{T-T_0}{2}$
  is the same. In the case of the three-level system, it is such that
  $\sigma_{\beta p}^2=2.0$ while in case of the many-level system
  $\sigma_{tot}^2=0.2$.}
\label{fig1}
\end{figure}

\subsection{Three-level system}
First a simple three-level system is considered. In this case a
full iterative solution for the optimizing driving frequency is
easy to calculate from Eq. (\ref{Delta solution}) (with complex
contributions neglected). System parameters have the following
values ($a.u.\equiv atomic \;units$): $\omega_{\beta
\alpha}=0.017671 \;a.u.$, $s_{\beta \alpha}=1$, $\mu_{\beta
\alpha}=0.073 \;a.u.$; $\omega_{\beta p}=0.017611 \;a.u.$,
$s_{\beta p}=-1$, $\mu_{\beta p}=0.098\; a.u.$. These system
parameters correspond to the three ro-vibrational levels of the HF
molecule in the ground electronic state: $\alpha \equiv
(v=0,j=2,m=0)$, $\beta \equiv (v=1,j=1,m=0)$, $p \equiv
(v=2,j=2,m=0)$. System parameters are such that the optimizing
frequency shift is away from the perturbing line. In all cases,
the total pulse duration $T-T_0$ equals 7.25 ns.

In order to present clearly the improvement that optimization of
driving frequency induces in population transfer between the two
selected levels, the perturbation strength in following examples
is set to an extreme value: $\sigma_{\beta p}^2=2.0$. Fig. 2
compares evolution of optimized frequency $\omega(t)$ with two
resonant frequencies of the system, $\omega_{\beta \alpha}$ and
$\omega_{\beta p}$. In Fig. 3 resonant and optimized population
dynamics are shown for each of envelope shapes. The increase in
the amplitude of the population transfer between the selected two
levels is obvious.

\begin{figure}
  \includegraphics[width=10cm]{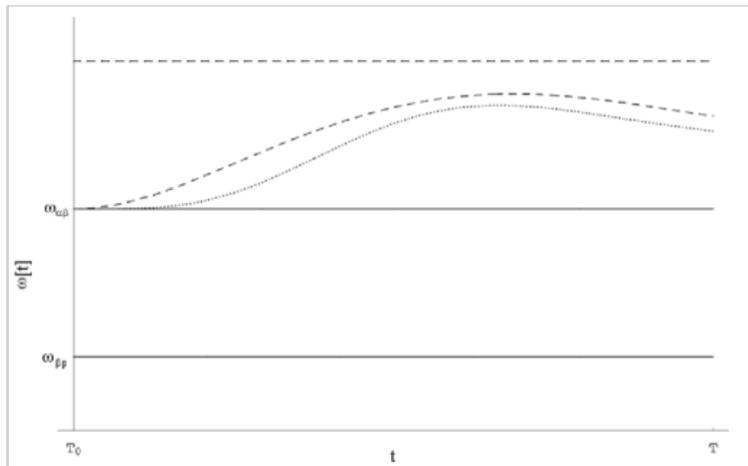}\\
  \caption{Optimized frequency plots for the three-level case. Time
  t is on the abscissa, the perturbation frequency $\omega(t)$ on the
  ordinate. The total pulse duration is $7.25 \; ns$. Two straight solid
  lines indicate the two resonant frequencies of the system,
  $\omega_{\beta \alpha}$ (upper) and $\omega_{\beta p}$ (lower).
  The remaining three lines are optimized frequencies for the three
  types of pulse: the dash-dotted for a square pulse, the dashed line for
  a sine pulse and the dotted for a sine squared pulse.}
\label{fig2}
\end{figure}

\begin{figure}
  \includegraphics[width=10cm]{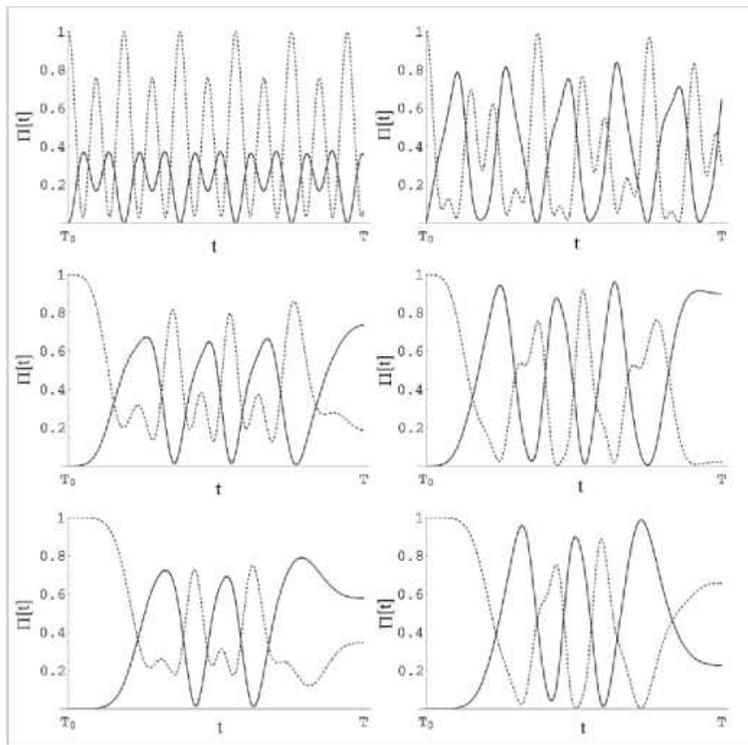}\\
  \caption{Comparison of the resonant and optimized population dynamics.
  Graphs on the left side present numerical solution to
  system dynamics for each of pulse types with a resonant
  perturbation applied, $\omega(t)=\omega_{\beta \alpha}$. Graphs on
  the right hand-side present optimized dynamics. The top row
  corresponds to a square pulse, the middle to a sine pulse and
  the bottom row to a sine square pulse. For the sake of clarity, only
  $\alpha$ (solid line) and $\beta$ (dashed line) populations
  are plotted while p population is omitted. Although the
  optimization clearly does not produce clean two-level dynamics,
  the increase in the amplitude of population oscillation is
  nevertheless evident.}
\label{fig3}
\end{figure}

\subsection{Many-level system}

As an example of a many-level system, the set of ro-vibrational
states of the HF molecule in the electronic ground state is
considered. The numerical model used for the calculation of the
system dynamics includes 310 levels (31 rotational $\times$ 10
vibrational). It is based on the HF internuclear potential data
and electric dipole moment data from \cite{mueller1998} and
\cite{zemke1991} respectively. The targeted transition is
$(v=1,j=1,m=0) \rightarrow (v=0,j=0,m=0)$). In all cases, the
total pulse duration $T-T_0$ equals 4.84 ns.

It was demonstrated in the previous subsection that optimization
indeed leads to improvement of the population transfer dynamics,
even when the perturbation is very large. However, as in such
conditions complete population transfer is unattainable, these
examples were more of a qualitative nature from the standpoint of
population transfer control. The many-level system considered now
is more realistic than the previous three-level one and the focus
is shifted to quantitative predictions. Hence, the employed drive
intensity will be much smaller so that results can be directly
applied to population transfer control. Pulse envelope shapes are
the same as in the three-level case (see Fig. 1). Maximum
amplitudes of electric field are likewise equal in all three
cases, but now they are chosen so that $\sigma_{tot}^2=0.2$. Since
the perturbation is relatively small, the optimized frequency may
be determined from the first-order approximate solution (\ref{bez
rot multi}).

\begin{figure}
  \includegraphics[width=10cm]{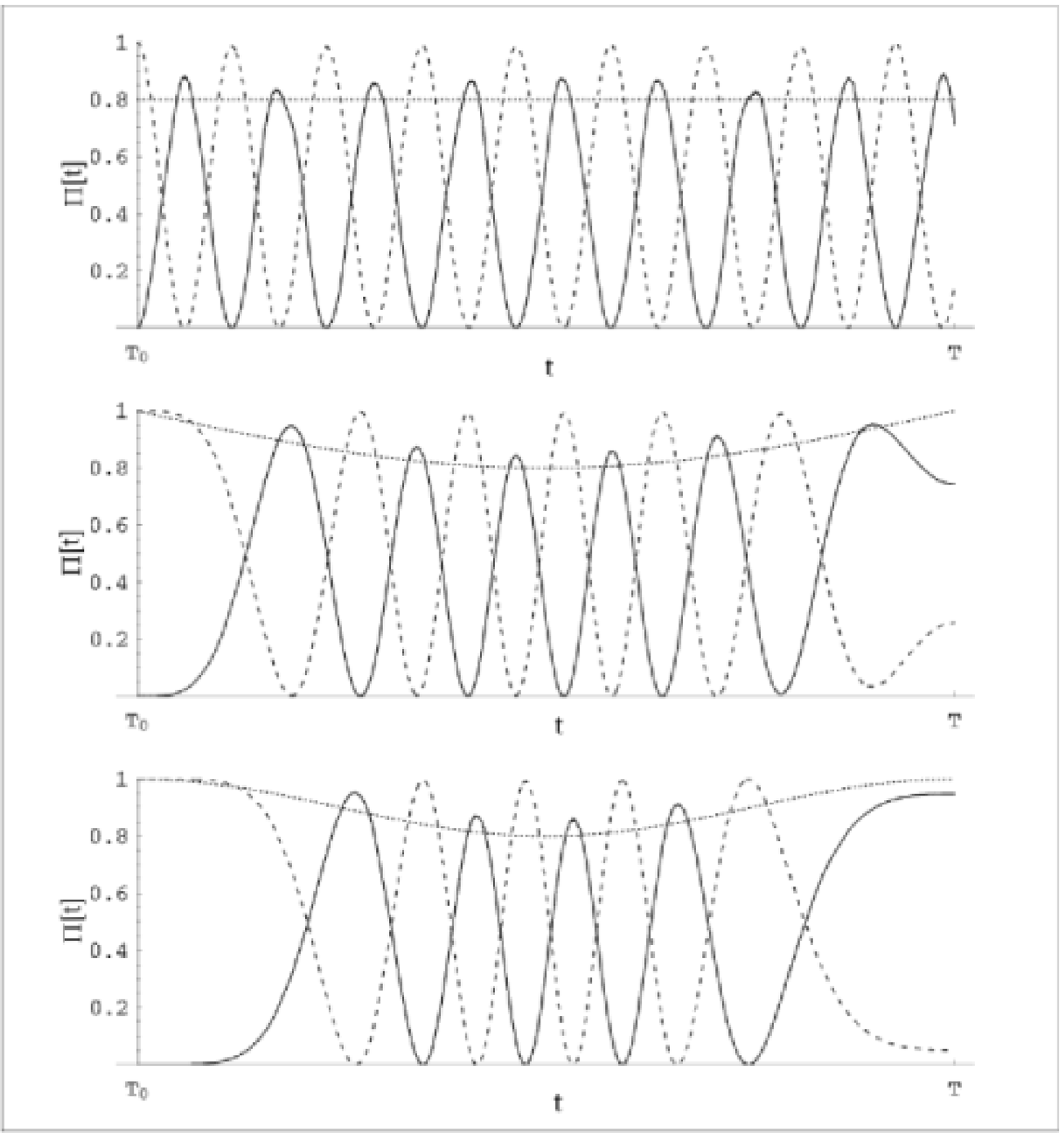} \\
  \caption{Optimized population dynamics of the two targeted levels
  in the many-level system. Time t is on the abscissa, populations
  $\Pi_{\alpha,\beta}(t)$ are on the ordinate. The top graph corresponds
  to a square pulse, the middle to a sine pulse and the bottom one
  to a sine squared pulse. In all cases $\sigma_{tot}^2=0.2$ and
  the total duration of the pulse is 4.84 ns. The dotted line on each
  graph indicates the function $1-m(t)\sigma_{tot}^2$.}
\label{fig4}
\end{figure}

In Fig. 4 the optimized dynamics of two target system levels is
shown for each of three pulse envelopes. In all cases two things
should be noted. First, the general shape of the optimized
dynamics of each of the two selected levels is fairly close to
pure sinusoidal oscillations. This is more so, the smaller the
perturbation strength parameter $\sigma_{tot}^2$ is. However, the
complete population transfer is again not achieved because a
certain share of the population unavoidably ends up in perturbing
levels. Second, the actual instantaneous loss of population
transfer is close to (and actually smaller than)
$m(t)\sigma_{tot}^2$, as indicated by dotted line in each of the
plots. This shows that $\sigma_{tot}^2$ indeed is a good
quantitative (and not just qualitative) indicator of relative
drive strength. The argument which led to relation (\ref{max
drive}) may hence again be applied to determine the maximum
intensity of the driving radiation to be employed if population
losses to perturbing levels are to be smaller than some
pre-defined amount.

\section{Conclusion}

As was stated in the introduction, the aim of this paper is to
explore and refine the use of Rabi oscillations as a tool in
selective population manipulation of complex discrete-level
quantum systems. The main aims of such manipulation are as great
as possible population transfer and at the same time as short as
possible population transfer time. From the simple two-level
theory it is well known that the an increase in drive intensity
yields a reduction in population oscillation period. However, the
same theory can neither fully disclose all the limitations of this
result that arise from the complexity of the internal structure of
a many-level system, nor can it handle the unavoidable loss of
population to the rest of the system. Results presented in this
paper fill this gap: they enable one to determine the maximum
possible drive intensity (and hence the lower limit of time) with
which oscillations of pre-selected amplitude (say 99\%) may be
achieved, and at the same time to minimize unavoidable losses of
the population to non-targeted system levels. Finally, the method
of Rabi spectra (see \cite {bonacci2003.1}) presents a simple, yet
useful conceptual supplement to the analysis presented in this
paper.

In order to achieve the quickest possible population transfer
between two pre-selected levels, driving pulse should be tailored
so that it produces only a single half-oscillation of the
population. However, during research for this paper it has been
noted that for strong fields standard $\pi$-pulse theory (see
equation (\ref{pi pulse}) and reference \cite{holthaus1994}) is
also deficient when it comes to the complex many-level systems.
Work is currently in progress on analytical extension of standard
$\pi$-pulse theory that would resolve this issue.

\section*{Acknowledgment}
I am very grateful to Dr. Nadja Do\v{s}li\'c for insightful
discussions and assistance during work on problems explored in
this paper. I am also grateful to Dr. Danko Bosanac for providing
the initial idea from which the topic of this research developed.


\end{document}